\documentclass{article}
\usepackage{spconf,amsmath,graphicx}
\usepackage{lipsum}
\usepackage{booktabs}
\usepackage{caption}
\usepackage{subcaption}
\usepackage{url}
\title{Cross-speaker Emotion Transfer by Manipulating Speech Style Latents}
\name{Suhee Jo, Younggun Lee, Yookyung Shin, Yeongtae Hwang, Taesu Kim}
\address{Neosapience, Inc.}
\begin{document}
\ninept
\maketitle

\begin{abstract}
In recent years, emotional text-to-speech has shown considerable progress. However, it requires a large amount of labeled data, which is not easily accessible. Even if it is possible to acquire an emotional speech dataset, there is still a limitation in controlling emotion intensity. In this work, we propose a novel method for cross-speaker emotion transfer and manipulation using vector arithmetic in latent style space. By leveraging only a few labeled samples, we generate emotional speech from reading-style speech without losing the speaker identity. Furthermore, emotion strength is readily controllable using a scalar value, providing an intuitive way for users to manipulate speech. Experimental results show the proposed method affords superior performance in terms of expressiveness, naturalness, and controllability, preserving speaker identity.

\end{abstract}
\section{Introduction}
\label{sec:intro}

As there are growing expectations for human-like TTS, subtle changes in prosody or emotion should be reflected in the output of a TTS model. Humans can speak with different emotions and this leads to rich and diverse conversations. However, emotional speech data are not easy to acquire. It is extremely hard to record multiple sentences for a long time while consistently preserving emotion. Moreover, due to the ambiguity of emotion labels, samples with inconsistent emotion labels are easily observed in open source emotional speech datasets \cite{cui2021emovie}. Even if it is possible to find a correct emotion label, expressiveness of emotion is limited without controllability of emotion intensity.

Most of the previous methods for emotional speech synthesis use additional emotion encoding or a reference audio \cite{lee2019robust,wu2019end,lei2022msemotts}. These models require an emotion label for each sample, consuming a fair amount of emotion-tagged speech data \cite{lee2019robust,lei2022msemotts}. Furthermore, inconsistency in emotion labels leads to degraded performance of emotional speech synthesis. Also, cross-speaker emotion transfer often does not work when an unseen emotion is transferred to a speaker. In regard to emotion strength control, there have been several attempts \cite{li2022cross, lei2021fine} to generate emotion strength scores using an emotion classifier or a ranking function. These approaches are still exposed to mislabeling problems, however, as they use emotion labeled data to extract emotion intensity.

To address the issues noted above, we propose transferable and controllable emotion speech synthesis by leveraging rich latent representation. Domain adversarial training and cycle-consistency loss disentangle the speaker from style, making the latent style space rich and transferable. During training, the entire model is trained without any emotion label. To transfer emotion, we utilize a SVM hyperplane during inference to manipulate the style latent towards a desirable emotion. Our method successfully transfers emotion to an emotion-neutral reading-style speaker with only a handful of emotion labeled samples. Furthermore, without external labels, emotion intensity can be controlled by a scalar value, which is easy and intuitive. The generated audio samples are available at \url{http://emo-transfer.github.io}

\section{Related Works}
\subsection{Emotional Speech Synthesis}
For emotional speech synthesis, it is common to use an emotion label as a global condition \cite{lee2019robust,wu2019end}. Otherwise, emotion information is extracted from a reference audio \cite{skerry2018towards,lei2022msemotts} or text \cite{shin2022text}. Most of these methods not only require a large quantity of emotion labeling, but also often fail to achieve good quality in cross-speaker emotion transfer. Some approaches \cite{cai2021emotion,zhou2021seen} use Speech Emotion Recognition (SER) to obtain emotion labels. However, SER is another challenging task and it still requires emotion tagged data to train itself. To reflect not only a type of emotion but also its intensity, methods for controlling emotion strength have been suggested \cite{li2022cross,lei2021fine,im2022emoq}. \cite{schnell2021improving} uses an external SER model to extract emotion intensity scores, whereas \cite{lei2021fine} uses a ranking function to predict emotion intensity.  However, for these models all data should still be labeled for training.

\subsection{Latent Space Manipulation}
In the image synthesis domain, attribute editing using StyleGAN \cite{karras2019style} has been widely studied  \cite{shen2020interfacegan, collins2020editing, abdal2021styleflow}. Such methods utilize latent space of StyleGAN to manipulate attributes. Among these methods, \cite{shen2020interfacegan} provides a simple yet effective approach for editing. Using a hyperplane in latent space that discriminates attribute-positive and attribute-negative samples, facial attributes such as age, gender or pose can be edited. The hyperplane is acquired from SVM training. In practice, any binary attributes can be manipulated when latent vectors from both positive and negative sides are given. In this paper, we adopt the method of latent space manipulation suggested in \cite{shen2020interfacegan} to synthesize emotional speech from emotion-neutral speakers. Details of training the SVM and manipulating latents will be described in Section \ref{Controllable Cross-speaker Emotion Transfer}

\section{Method}
\label{sec:method}
To facilitate cross-speaker emotion transfer, we propose domain adversarial training and cycle-consistency loss for the acoustic model to learn disentangled style and speaker representation. Along with this model, we suggest a method for controlling emotion and its intensity by utilizing a hyperplane obtained from training a SVM.

\begin{figure*}[t]
  \centering
     \begin{subfigure}[b]{0.6\textwidth}
         \centering
         \includegraphics[width=\textwidth]{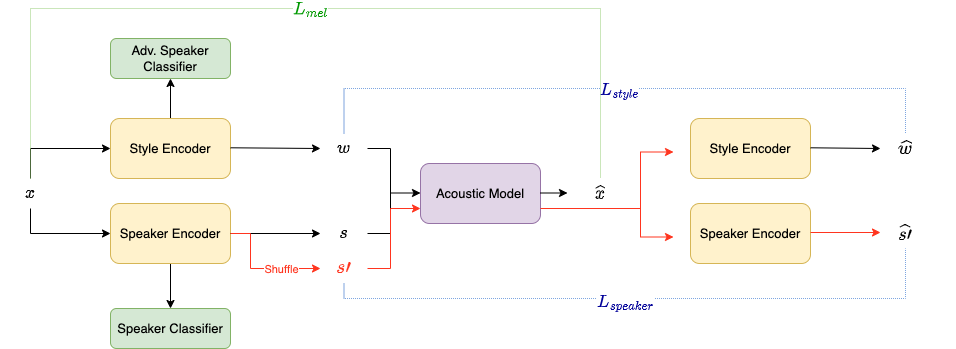}
         \caption{The overall architecture of the proposed acoustic model}
         \label{fig:overall}
     \end{subfigure}
    \hspace{0.4pt}
     \begin{subfigure}[b]{0.3\textwidth}
         \centering
         \includegraphics[width=\textwidth]{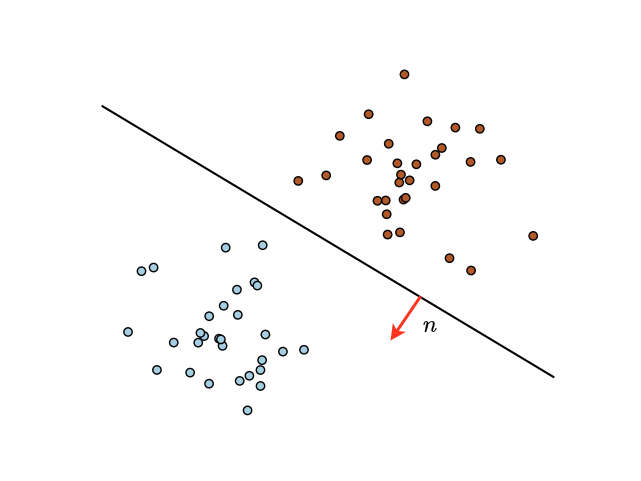}
         \caption{SVM hyperplane}
         \label{fig:svm}
     \end{subfigure}
    \caption{From the first stage (\subref{fig:overall}), speaker-independent style vectors are extracted. In the second stage (\subref{fig:svm}), we use a unit vector \(n\) to transfer emotion. The unit vector is perpendicular to a SVM hyperplane that separates emotional style vectors from emotion-neutral ones.}
    \label{fig:model}
\end{figure*}

\subsection{Disentangled Latent Style Vectors}
We focus on extracting rich yet disentangled style representation from speech. If style space is disentangled from speaker identity, cross-speaker emotion transfer becomes easier. Otherwise, latent vector manipulation using a SVM will not work well, as a direction vector acquired from the SVM will also transform speaker information. For example, converting an emotion-neutral sample to an angry one can result in a change of the speaker identity. Therefore, we try to disentangle the speaker identity to generate better style latent space for emotion transfer.

\subsubsection{Tacotron2-based Acoustic Model}
Our proposed method is based on Tacotron2 with the following modifications. The main difference is that we adopt a style encoder and a speaker encoder. The style encoder has a target mel spectrogram as an input and generates a style vector. The architecture of the style encoder is based on \cite{min2021meta}. The style encoder generates the final output, called a style vector, which is added to the output of a text encoder. The speaker encoder also receives a mel spectrogram as an input and generates a speaker vector. The speaker encoder is composed of a LSTM and a projection layer. The speaker vector is concatenated to an input of a decoder. 

\subsubsection{Domain Adversarial Training}
By adversarially classifying a style vector to a speaker class, the style encoder learns speaker-independent style. In the speech domain, there have been many applications of domain adversarial training, such as \cite{im2022emoq, zhang2019learning}. This has been shown to be largely effective in disentangling information. Our adversarial speaker classifier consists of linear layers. Prior to the layers, a gradient reversal layer was attached. The gradient reversal layer reverses gradients from the classifier so that a style vector cannot discriminate speakers. On the speaker encoder side, a speaker classifier follows a speaker encoder to maintain speaker information. The speaker classifier shares the same structure with the adversarial speaker classifier. The only difference is that it does not have a gradient reversal layer.

\subsubsection{Cycle-consistency Loss}
For both the style encoder and speaker encoder, we adopt cycle-consistency loss to preserve information. If we let the speaker encoder \(E_s\) and the style encoder \(E_w\) and the target mel spectrogram as \(x\), a style vector \(s\) and a speaker vector \(w\) are the outputs of  \(E_s({x})\) and \(E_w({x})\), respectively. The generated output of the entire TTS model is \(  f(t, w, s) \), where \(t\), \(w\), \(s\) are the input text, style vector, and speaker vector, respectively. After generating a predicted output \(\hat{x}\),  we use speaker vectors from randomly sampled speakers to produce  \( {s}' \). Then, \( {x}' \) is generated by \( f(t, w, {s}') \).  \( {x}'\) is used as an input of both the speaker encoder and the style encoder, which leads to \(E_s({x}')\), \(E_w({x}')\). Cycle-consistency loss for both the speaker and style can be written as below. \(N\) is the batch size during training. We use Mean Squared Error (MSE) for the loss term. 
\begin{equation} \label{eq:1}
L_{style} = \frac{1}{N}\sum_{i=1}^{N} (w_{i} - E_w({x_i}')))^2
\end{equation}
\begin{equation} \label{eq:2}
L_{speaker} = \frac{1}{N}\sum_{i=1}^{N} ({s_{i}'} - E_s({x_i}')))^2
\end{equation}

\subsection{Controllable Cross-speaker Emotion Transfer} \label{Controllable Cross-speaker Emotion Transfer}
Before inference, we extract speaker vectors from all utterances of a target speaker. We then obtain the centroid of those vectors to use it as a speaker vector for inference. Likewise, a style vector that is fed to the model during inference is a centroid of all style vectors from the target speaker. A style vector without any manipulation represents emotion-neutral style, as the speakers we experimented on are emotion-neutral speakers. To transfer emotion, we edit the style vector using a SVM. 

We train a SVM with a positive group of style vectors from a certain emotion and a negative group from emotion-neutral samples. As \cite{arik2018neural} showed gender and accent transformation through vector operations, it can be assumed that there exists a hyperplane for each attribute that separates attribute-positive and attribute-negative latent vectors. To find a hyperplane, we use a linear SVM. After training the SVM, we define a unit normal vector \(n\) that is perpendicular to the hyperplane with its distance to the hyperplane 0. By using \(n\), we can determine on which side of the hyperplane a style vector \(w\) lies. We use a metric defined as \( d(n,w) = n^{T}w \). If \( d(n,w)\) is negative,  \(w\) is on the attribute-negative side. On the other hand, If \( d(n,w)\) is positive,  \(w\) is attribute-positive. 
In our case, \(w\) is the centroid of all style vectors from a target speaker, as described above. By adding \(n\) multiplied by a scaling factor \(\alpha\) to \(w\), as in (\ref{eq:3}), emotion can be adjusted. 

\begin{equation} \label{eq:3}
w_{edit} = w + \alpha n
\end{equation}

If \(\alpha > 0 \), the emotion of the edited vector \(w_{edit}\) is led towards the direction of an intended emotion since \( d(n,w_{edit}) \) becomes \( d(n,w) + \alpha\). If \(\alpha < 0 \), \(w_{edit}\) moves towards the opposite direction. By increasing \(\alpha\) ranging from [0, 2], a gradual change in emotion intensity is shown. One more interesting point is that even if \(\alpha\) is set to a negative value, it gives a meaningful result. For example, if there exists an editing vector that manipulates a style vector towards sad emotion, we can adjust \(\alpha\) to a negative value to make it sound happy.

To manipulate emotion-related style alone without affecting other speaking style from a speaker, we use conditional manipulation. Given \(n_{1}\) and \(n_{2}\) from two hyperplanes, we obtain a projected direction as \(n_{1} -(n_{1}^{T}n_{2})n_{2}\) that makes  \(n_{1}\) independent of the direction \(n_{2}\). As we are trying to manipulate style, \(n_{1}\) is a perpendicular unit normal vector from a hyperplane learned from an emotion classification and \(n_{2}\) will be from a speaker classification. 

\begin{table}[t]
    \centering
    \resizebox{0.8\columnwidth}{!}{%
    \begin{tabular}{c | c || c | c }
    \noalign{\smallskip}\noalign{\smallskip}\noalign{\smallskip}\noalign{\smallskip}\hline\hline
    \cline{1-4}
        \textbf{Style}  & \textbf{Ratio [\%]} & \textbf{Style}  & \textbf{Ratio[\%]} \\
    \hline
    \hline
     Reading & 60.4 & Fairytale & 1.3  \\
     \hline
     Conversational & 1.2 & Angry & 2.1  \\
     \hline
     Animation Dubbing & 12.1 & Sad & 2.0  \\
     \hline
     Whisper & 1.0 & Happy & 2.3 \\
     \hline
     Children & 7.6 & Not defined & 10.0 \\
    \hline
    \hline
    \end{tabular}%
}
\caption{Speech Style of the Training Data}
\label{table1}
\end{table}

\section{Experiments and Results}
\subsection{Dataset}
In this paper, we conducted experiments in both English and Korean. Our English dataset consists of 25 female speakers and 42 male speakers, with 127 hours in total, whereas our Korean dataset consists of 58 female speakers and 42 male speakers, with 270 hours in total. While an open-source dataset was also used, most of the samples were collected internally, recorded by professional voice actors. For open-source dataset, we used DailyTalk \cite{lee2022dailytalk} and Korean Single Speaker (KSS) Speech Dataset \cite{kss_dataset}. Table \ref{table1} shows the training data and its speech style used for the acoustic model. 

\begin{figure}[t]
  \centering
  \includegraphics[width=0.65\columnwidth]{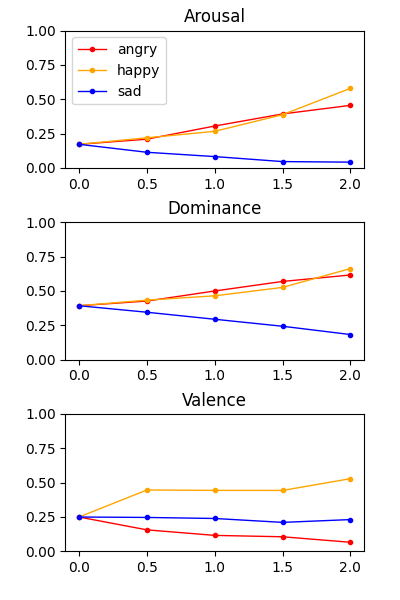}
  \caption{VAD (Valence-Arousal-Dominance) plot according to emotion and its intensity.}
\label{fig:vad}
\end{figure}

\subsection{Experimental Setup}
\label{sec:exp_setup}
First, for training the acoustic mode, we follow details of the style encoder of \cite{min2021meta}. The speaker encoder consists of one LSTM layer with hidden size of 258 and one projection layer. Both the style vector and speaker vector are 512 dimensional. Both the adversarial speaker classifier and speaker classifier consist of three fully-connected layers, followed by ReLU activation and dropout rate by 0.1 for each layer. All the classification losses are multiplied by 0.02 and added to loss terms of the acoustic model. In the training process, we train the model with batch size of 32 using the Adam optimizer. Learning rate scheduling follows the Noam Scheduler, with initial learning rate \(10^{-3}\) and a warm-up threshold of 4000 steps. We trained the model for each gender and language, resulting in four models. All the models were trained for 60 epochs. Until the attention loss reaches 0.7, the attention module is trained alone to stabilize the training process.

To train the SVM, style latent vectors for a negative set and a positive set were collected. The negative set consists of emotion-neutral samples, spoken in reading-style. The positive set consists of emotional samples of a desirable emotion, among angry, happy, and sad. Normally, 100 samples were randomly selected for each negative and positive set among training samples that we could find labels of. Although using 100 samples showed stable performance, results on a single sample were good as well. When we train a SVM under a one-shot setting, we selected paired samples, which share the same script and speaker to eliminate confounding variables other than emotion. In the classification task, our SVM models achieved over 90\%  accuracy on a validation set. Additionally, a speaker classification task was conducted for conditional manipulation. We used all samples from the positive speaker and randomly selected the same number of negative samples. The speaker classification task achieved over 98\%  accuracy on a validation set.

\begin{table}[t]
    \large
    \centering
    \resizebox{\columnwidth}{!}{
    \begin{tabular}{c | c | c | c}
    \noalign{\smallskip}\noalign{\smallskip}\hline\hline
    \cline{1-4}
        \textbf{Setting}  & \textbf{Naturalness}  & \textbf{Emo. Similarity} & \textbf{Spk. Similarity} \\
    \hline
     Ground Truth & $4.55 \pm 0.05$ & - & - \\
     Baseline & $3.54 \pm 0.08$ & $2.48 \pm 0.08$ & $1.90 \pm 0.07$ \\
     Conventional \cite{li2022cross} & $3.86 \pm 0.04$ & $2.91 \pm 0.04$ & $2.46 \pm 0.05$ \\
     \hline
     \textbf{Proposed} & $\textbf{4.70} \pm \textbf{0.04}$ & $\textbf{4.24} \pm \textbf{0.06}$ & $\textbf{3.98} \pm \textbf{0.07}$ \\
    \hline
    \hline
    \end{tabular}
    }
    \caption{MOS on naturalness, speaker similarity and emotion similarity}
    \label{tab:table2}
\end{table}

\subsection{Evaluation}
We conducted a subjective evaluation using mean opinion score (MOS) to evaluate naturalness, speaker similarity, and emotion similarity. In the test, 15 subjects were asked to rate 180 sentences on a scale from 1 to 5. The generated samples were randomly selected from each emotion category. The number of samples per each category was balanced. For speaker similarity, a ground truth sample was given for each entry, as a reference. The participants compared a given sample with the ground truth on the basis of speaker identity. For emotion similarity, an emotion tag, for example, "happy", was given and the subjects were asked to judge whether a given speech sample expresses the emotion. To ease their decision, we provided a corresponding emotion-neutral sample that was generated from the sample model with the same text but without emotion transfer. For the baseline, we used Tacotron2. The baseline model uses speaker embedding to encode speaker information while maintaining a style encoder which has identical structure with the proposed model. The baseline generates output with an averaged style vector, which was derived from 100 randomly selected emotion-labeled style vectors for each emotion. We used a HiFi-GAN \cite{kong2020hifi} as a vocoder to generate waveforms.

Table \ref{tab:table2} shows the results of the evaluation of the proposed model along with ground truth, the baseline and the conventional model \cite{li2022cross}. With respect to the MOS score, the proposed model outperforms the baseline model and the conventional model in all evaluation categories. In terms of naturalness, the proposed model was rated higher than the ground truth. As we collected speech data with various styles regardless of the source, a few ground truth samples contain unnatural prosody or noise. However, during training, this phenomenon appears to be diluted. We also observed that the baseline model and the conventional model fall behind the proposed model due to its vulnerability to a wide range of variation in style. Naturally, emotion similarity and speaker similarity are at variance with each other because speech features such as pitch or timber vary greatly from original speech samples as emotion becomes intense. Even with this contradiction, the proposed model shows high scores in both emotion similarity and speaker similarity. Overall, the proposed model maintains high fidelity in speaker identity and naturalness, expressing proper emotion at the same time.

\begin{figure}[t]
  \centering
  \includegraphics[width=0.9\columnwidth]{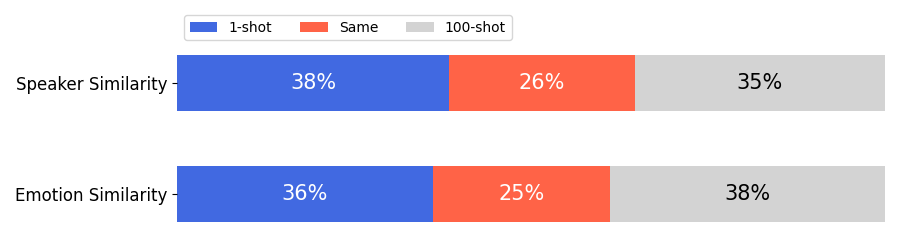}
  \caption{AB preference test on emotion similarity and speaker similarity}
\label{fig:preference}
\end{figure}

\subsubsection{Emotion Intensity Control}
To test emotion intensity control, we used a pretrained speech emotion recognition (SER) model \cite{wagner_johannes_2022_6221127}. This SER model was trained using MSP-Podcast \cite{lotfian2017building}. We used the pretrained model as it was without any fine-tuning. MSP-Podcast was labeled with arousal, valence, and dominance ranging from 1 to 7. \cite{wagner_johannes_2022_6221127} normalized the values into an interval of 0 to 1. Happiness and anger are known to include high degrees of arousal and dominance whereas sadness has low degrees of arousal and dominance. Happiness also has high degrees of valence while anger and sadness are low in valence.

In Figure \ref{fig:vad}, the x axis represents \(\alpha\), a scaling factor for emotion intensity, and the y axis represents prediction for arousal, valence, and dominance ranging from 0 to 1. Fifty randomly selected  samples were used to extract a mean value for each metric. As shown in Figure \ref{fig:vad}, both for angry and happy, arousal increases as we raise the intensity of emotion by controlling \(\alpha\), whereas arousal decreases for sad. Valence increases for happy whereas it slightly decreases for both angry and sad emotion. The overall tendency shows that emotion intensity is well adjusted towards an intended emotion. 

\subsubsection{Few-shot Emotion Transfer}
To demonstrate few-shot emotion transfer, we compare one-shot emotion transfer to 100-shot setting in an A/B preference test. In the preference test, participants were asked to select which of the two samples is more similar to a given speaker or emotion. As shown in Figure \ref{fig:preference}, one-shot setting shows comparable results to 100-shot. Neither showed superior results in terms of preference.

\subsubsection{Ablation Study}
\begin{figure}[t]
  \centering
  \includegraphics[width=0.6\columnwidth]{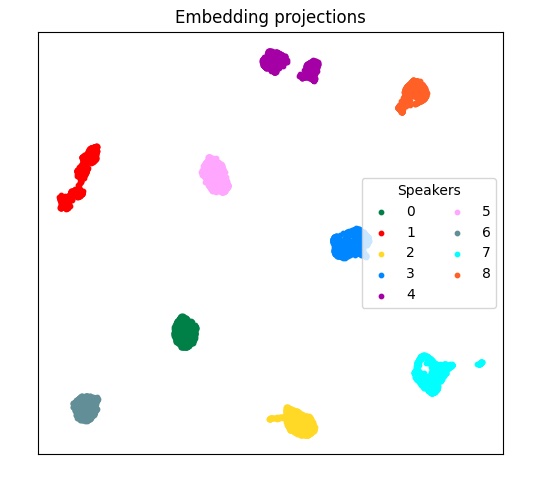}
  \caption{Visualization of speaker vectors learned by a speaker encoder.}
\label{fig:embeddings}
\end{figure}

To show the effectiveness of each module of the proposed model, we conducted an ablation study. Table \ref{table3} represents MOS scores for each case in terms of naturalness, speaker similarity and emotion similarity. Removing an adversarial speaker classifier or cycle-consistency loss resulted in degradation in emotion similarity. In particular, the score for emotion similarity drops drastically without cycle-consistency loss. It can be inferred that cycle-consistency loss plays an important role in generating disentangled yet rich latent style space. Even though it was highly rated in speaker similarity, this is because emotion was not transferred. In comparison, the proposed model shows superior performance in terms of emotion expression, maintaining sufficiently good scores in speaker similarity. This indicates that both components are effective in disentangling and preserving style and speaker information. Figure \ref{fig:embeddings} also supports this argument, showing that speaker vectors learned by the model successfully reserve speaker identity.  

\begin{table}[t]
    \Huge
    
    \centering
    \resizebox{\columnwidth}{!}{
    \begin{tabular}{c | c | c | c}
    \noalign{\smallskip}\noalign{\smallskip}\hline\hline
    \cline{1-4}
          \textbf{Setting}  & \textbf{Naturalness}  & \textbf{Emo. Similarity} & \textbf{Spk. Similarity}   \\
    \hline
     \textbf{Proposed} &  $\textbf{4.77} \pm \textbf{0.04}$ & $\textbf{4.32} \pm \textbf{0.07}$ & $\textbf{4.31} \pm \textbf{0.09}$ \\
     \hline
     w/o Adv. Speaker Classifier  &   $4.72 \pm 0.04$ & $4.13 \pm 0.07$ & $4.34 \pm 0.06$ \\
     w/o Cycle-consistency loss  &   $4.77 \pm 0.04$ & $3.55 \pm 0.08$ & $4.71 \pm 0.05$  \\  
    \hline
    \hline
    \end{tabular}
    }
    \caption{Ablation Study}
    \label{table3}
\end{table}

\section{Conclusion}
This paper suggests a novel method for cross-speaker emotion transfer and manipulation by applying vector arithmetic on a disentangled latent style space. To extract latent style without interference from speaker information, we propose domain adversarial training and cycle-consistency loss. In addition, we provide an intuitive way to transfer and to manipulate the style latent vector by using a SVM hyperplane. Experimental results show that our method greatly improves speaker similarity and emotion similarity while keeping naturalness, without leveraging a large amount of emotion labeled data. In future work, we will conduct experiments on other semantic attributes of speech data, such as age or gender.

\vfill\pagebreak
\bibliographystyle{IEEEbib}
%\bibliography{strings,refs}
\bibliography{mybib}

\end{document}